\documentclass[12pt]{article}
\usepackage{amssymb}
\usepackage{color,graphicx}
\usepackage{amsmath}
\usepackage{hyperref}

\begin{document}

\renewcommand{\abstractname}{\hfill}

\newpage
\pagenumbering{arabic}
\begin{center}
\LARGE \bf {
Static limit and Penrose effect in\\ rotating reference frames}
\end{center}

\begin{center}
\large \bf
A. A. Grib\footnote{\,Herzen  State Pedagogical University, Saint Petersburg,
Russia;
A. Friedmann Laboratory for Theoretical Physics, Saint Petersburg, Russia,
e-mail:\, andrei\_grib@mail.ru} and
Yu. V. Pavlov\footnote{\,Institute of Problems in Mechanical Engineering,
Russian Academy of Sciences, Saint Petersburg, Russia;
N.I.\,Lobachevsky Institute of Mathematics and Mechanics,
Kazan Federal University, Kazan, Russia, e-mail:\, yuri.pavlov@mail.ru}
\end{center}

\begin{abstract}
    We show that effects similar to those for a rotating black hole arise for
an observer using a uniformly rotating reference frame in a flat space-time:
a surface appears such that no body can be stationary beyond this surface,
while the particle energy can be either zero or negative.
    Beyond this surface, which is similar to the static limit for a rotating
black hole, an effect similar to the Penrose effect is possible.
We consider the example where one of the fragments of a particle that has
decayed into two particles beyond the static limit flies into the rotating
reference frame inside the static limit and has an energy greater than the
original particle energy.
    We obtain constraints on the relative velocity of the decay products
during the Penrose process in the rotating reference frame.
   We consider the problem of defining energy in a noninertial reference frame.
    For a uniformly rotating reference frame, we consider the states of
particles with minimum energy and show the relation of this quantity to the
radiation frequency shift of the rotating body due to
the transverse Doppler effect.
\end{abstract}


{\centering \section{\large Introduction}}

    The study of relativistic effects related to rotation began immediately
after the theory of relativity appeared~\cite{Ehrenfest1909},
and such effects are still actively discussed in
the literature~\cite{EhrenfestVrash}.
    The importance of studying
such effects is obvious in relation to the Earth's rotation.
    In~\cite{GribPavlov2016d}, we compared the properties of particles
in rotating reference frames in a flat space with the properties of
particles in the metric of a rotating black hole.
    We showed that similarly to the metric of the Kerr black hole,
there is a surface (static limit) in a uniformly rotating reference
frame in a flat space, beyond which no body can be in the rest state
and particle states with negative energy are possible, as in
the black hole ergosphere~\cite{MTW}.
    The spatial infinity of a rotating reference frame plays
the role of the event horizon in a rotating black hole.

    The existence of a region where the particle energy can be negative
poses a problem of the possible existence of an effect analogous to one
proposed by Penrose for rotating black holes where one of the
fragments in particle decay in the ergosphere can have an energy greater
than the original particle energy~\cite{Penrose69}.
    The Penrose effect allows extracting energy from a rotating black hole.

    At first glance, the occurrence of the Penrose effect in a rotating
reference frame would lead to an impossible process where the energy of one
of the fragments in decay in a flat space-time is greater than the
original decaying particle energy.
    This is obviously impossible if we mean the energy defined and
measured in the inertial reference frame in the Minkowski space-time.
    But the definition of energy depends on the chosen reference frame,
on the choice of the appropriate Killing vector.
    Here, we show that the standard definitions of energy in the case
of a uniformly rotating reference frame leads to the same expression for
the energy of a freely moving particle and that an effect similar to
the Penrose effect in black holes can be realized for the energy thus defined.

    The fact that the other definition of the energy in a noninertial
reference frame leads to new physics is known in connection with
the Unruh effect~\cite{Unruh76}.
    Of course, particle decay interpreted as the Penrose effect in
rotating coordinates is the usual decay in an inertial reference frame
without any negative or zero energy.
    Nevertheless, not just any usual decay can be treated as the
Penrose effect.
    We obtain inequalities for the relative velocity of the fragments
that must be satisfied for such an interpretation, showing that the process
must be relativistic at least near the static limit.

\vspace{4mm}
{\centering \section{\large Rotating reference frames}}

    The interval in the Minkowski space written in cylindrical coordinates
$r'$, $\varphi'$, $z'$ is
    \begin{equation}    \label{v1}
d s^2 = c^2 dt^2 - d r^{\prime\,2} - r^{\prime\,2} d \varphi^{\prime\,2}
- d z^{\prime\,2}
\end{equation}
    We let $r$, $\varphi$, $z$ denote the rotating cylindrical coordinates and
assume that the rotation axis coincides with the coordinate axes $z$ and $z'$:
    \begin{equation}    \label{v2}
r'=r, \ \ \ \ z'=z, \ \ \ \ \varphi' = \varphi - \Omega t,
\end{equation}
    where $\Omega \ge 0$ is the angular rotation velocity.
        Substituting~(\ref{v2}) in~(\ref{v1}) yields expressions for the interval
and for the metric tensor in the rotating reference frame:
    \begin{equation}    \label{v3}
d s^2 = (c^2 - \Omega^2 r^2)\, dt^2 + 2 \Omega r^2 d \varphi\, d t -
d r^{2} - r^{2} d \varphi^{2} - d z^{2},
\end{equation}
  $$ 
\left( g_{ik} \right) =
\left( \begin{array}{crcr}
\displaystyle 1 - \frac{\Omega^2 r^2}{c^2} & 0
& \displaystyle \ \frac{\Omega r^2}{c} & 0 \\[2mm]
0 & - 1 & 0 & 0 \\[2mm]
\displaystyle  \frac{\Omega r^2}{c} & 0 &
- r^2 & 0 \\[2mm]
0 & 0 & 0 & -1
\end{array} \right), \ \
\left( g^{ik} \right) =
\left( \begin{array}{rrcr}
1 & 0 & \displaystyle \frac{\Omega}{c} & 0 \\[2mm]
0 & - 1 & 0 & 0 \\[2mm]
\displaystyle  \frac{\Omega}{c} & 0 &
\displaystyle \ \frac{\Omega^2}{c^2} - \frac{1}{r^2} & 0 \\[2mm]
0 & 0 & 0 & -1
\end{array}
\right),
$$ 
    $ i,k = 0,1,2,3 $.

    It was claimed in the well-known textbook~\cite{LL_II} that
``the rotating system of reference can be used only out to distance equal
to $ c / \Omega$.
        In fact, from~(\ref{v3}) we see that for $ r > c / \Omega$,
$g_{00}$ becomes negative, which is not admissible.
    The inapplicability of the rotating reference system at large distances is
related to the fact that there the velocity would become greater than
the velocity of light, and therefore such a system cannot be made up from
real bodies.''

    It was also noted in~\cite{Fok} that passing to a uniformly rotating
reference frame is possible only for distances from the rotation axis
not exceeding $ c / \Omega$.
    But the equivalence of inertial and rotating reference frames was
discussed in~\cite{EinsteinInfeld,Friedmann,BornETO}.

    We note that $ {\rm det}\! \left( g_{ik} \right) = - r^2$ and
metric~(\ref{v3}) is therefore nondegenerate for $r>0$,
although $g_{00}=0$ at $ r= c / \Omega$.
    The Earth's angular rotation velocity is equal to
${\Omega_\oplus \approx 7.29 \cdot 10^{-5}}$\,s$^{-1}$
(because the rotation period of the Earth relative to fixed stars,
the sidereal day, is 23 h 56 min 4 s \cite{FizikaKosmosa})
and the corresponding distance where $g_{00}=0$ is
$ c/\Omega_\oplus = 4.11 \cdot 10^9$\,km;
this distance is less then the Neptune orbit $ r \approx 4.5\cdot 10^9$\,km,
but greater than the Uranus orbit $ r \approx 2.9\cdot 10^9$\,km
(see~\cite{FizikaKosmosa}).

    Obviously, the reference frame related to the rotating Earth is used
practically in any laboratory, and astronomers used it not only over
distances less then the Neptune orbit but also to much larger
intergalactic scales up to the boundary of the observed Universe.
    Because body motion with the velocity exceeding the speed of light
is impossible, the rotating reference frame at distances from
the rotation axis greater than $ r= c / \Omega$ cannot be realized by
solid bodies.
    Nevertheless, the general covariance of general relativity theory
allows using of a rotating reference frame over all distances.

    Moreover, the analogy with the situation inside the ergosphere of
a rotating black hole shows that despite the negative sign of $g_{00}$,
the square of the interval for causally related points turns out to be
greater than zero, which is related to the presence of the nondiagonal
term $ g_{0 \varphi} d \varphi d t$.
    There is no real motion with
a velocity exceeding the speed of light, which is analogous to the known
``rotating spotlight'' paradox.

    The surface $ r= c / \Omega$ plays the role of the static limit for
a rotating black hole in the Boyer-Lindquist
coordinates~\cite{BoyerLindquist67}.
    It was shown in~\cite{GribPavlov2016d} that the particle energy
relative to the rotating reference frame can have negative values
in the region $ r > c / \Omega$.
    Below, we consider this phenomenon in detail and study
the possibility for observing the Penrose effect during decay of a particle
in the region $ r > c / \Omega$ into two fragments, the energy of one
of which is negative.

\vspace{4mm}
{\centering \section{\large Energy of a point particle in a space--time
of general form}
\label{secNewE}}

    The geodesic equations for a particle moving in a space-time with the
interval $ds^2 = g_{ik} dx^i dx^k$ can be obtained from
the Lagrangian~\cite{Chandrasekhar}
    $$ 
L = \frac{g_{ik}}{2}\, \frac{ d x^i}{d \lambda} \frac{ d x^k}{d \lambda},
$$ 
    where $\lambda$ is the affine parameter on the geodesic.
        For a timelike geodesic, $\lambda = \tau /m$,
where $\tau$ is the proper time of a moving particle of mass $m$.
    The generalized momentum can be defined as
    $$ 
p_i  \stackrel{\rm def}{=} \frac{\partial L}{\partial \dot{x}^i}
= g_{ik} \frac{d x^k}{d \lambda } ,
$$ 
    where $ \dot{x}^i = d x^i/d \lambda $.
    If the metric is stationary (its components are independent of the time~$t$),
then the zeroth covariant component of the momentum $p_0$ is conserved by virtue
of the Euler-Lagrange equation.
    The quantity $p_0$ for a massive particle equals the particle energy divided
by the speed of light~$E/c$:
    \begin{equation}    \label{dE1}
E=p_0 c = m c\, g_{0k} \frac{d x^k}{d \tau }.
\end{equation}
    Using the Killing vectors results in the same expression for the energy.
        If $\zeta$ is some Killing vector, then the quantity
    \begin{equation} \label{NEnergyEgen}
E^{(\zeta)} = m c^2\, \frac{dx^i}{ds}\,  g_{ik} \zeta^k =
m c^2 \, (u, \zeta) = c (p, \zeta)
\end{equation}
is conserved along the geodesic (see problem 10.10 in~\cite{LPPT}).
    Here $ u^i= dx^i / ds $ is the 4-velocity, and, \,
$ p^i= m\, c\, dx^i / ds $ is the particle 4-momentum.
    The value of quantity~(\ref{NEnergyEgen}) obviously depends on the particular
choice of the vector~$\zeta$.
    If the metric is time-independent and we choose $\zeta = (1,0,0,0)$, i.e.,
the Killing vector of a time-coordinate translation, then the value of $E^{(\zeta)}$
corresponds to the energy calculated by formula~(\ref{dE1}).

    In Cartesian coordinates of Minkowski space, the energy-momentum vector of
a free particle is
    $$ 
p'^{\, i} = \left( \frac{E'}{c}, \ {\bf p'} \right), \ \ \ \
p'_{\, i} = \left( \frac{E'}{c}, \ {\bf - p'} \right),
$$ 
    where ${\bf p'}$ is the usual 3-momentum (see~\cite{LL_II},~\S\,9).
    In cylindrical coordinates,
    \begin{equation}    \label{v6d}
p'^{\, i} = m \frac{d x^{\prime\,i}}{d \tau} = \left( \frac{E'}{c}, \
p^{\prime\,r}, \ p^{\prime \varphi}, \ p^{\prime z} \right), \ \ \
p'_{\, i} = \left( \frac{E'}{c}, \ - p^{\prime\,r},
\ - L'_z, \ - p^{\prime z} \right),
\end{equation}
    where
    $$ 
L'_z = r^2 p^{\prime \varphi} = m r^2 \frac{d \varphi' }{d \tau}
= \frac{E'}{c^2} r^2 \frac{d \varphi' }{d t}
$$ 
    is the projection of the angular momentum on the~$z$ axis.
        Because the components of the metric in cylindrical coordinates are
independent of $\varphi$, the covariant component $- L'_z$ of the momentum
is conserved.
    We note that the contravariant component $p^{\prime \varphi}$ is obviously
not conserved.

    We can obtain the energy-momentum vector in the rotating reference frame
by transforming~(\ref{v6d}) to the rotating coordinates:
    $$ 
p^{i} = \frac{ \partial x^i }{ \partial x^{\prime\,k} } p^{\prime\,k}
= \left( \frac{E'}{c}, \ p^{\prime\,r}, \ p^{\prime \varphi}
+ \Omega \frac{E'}{c^2}, \ p^{\prime z} \right), \ \ \
p_{i} = \left( \frac{E' + \Omega L'_z}{c},
\ - p^{\prime\,r}, \ - L'_z, \ - p^{\prime z} \right).
$$ 
    Therefore, the energy and projection of the angular momentum
(see equality~(\ref{dE1})) in a uniformly rotating reference frame are
    \begin{equation}    \label{v10v}
E = E' + \Omega L'_z,  \ \ \ \ L_z = L'_z.
\end{equation}
    As can be seen from~(\ref{v10v}) the energy in the rotating reference
frame differs from the energy in the nonrotating reference frame and can
even be negative!

    We here note that the obtained relativistic formula~(\ref{v10v})
coincides with the nonrelativistic expression in~\cite{LL_I}.
    We show that expression for the energy in~(\ref{v10v}) in
the nonrelativistic case can be obtained using the standard definition of
work as a scalar product force times translation.
    In a uniformly rotating reference frame with the angular
velocity $\mathbf{\Omega}$
    \begin{equation}    \label{ro1}
m \frac{d \mathbf{v}}{d t} = - \frac{\partial U}{ \partial \mathbf{r}} +
2 m \left[ \mathbf{v} , \mathbf{\Omega} \right] +
m \left[ \mathbf{\Omega} , \left[ \mathbf{v} , \mathbf{\Omega} \right] \right],
\end{equation}
    where $U$ is the potential energy of the particle
(see formula (39.9) in~\cite{LL_I}).
    Defining force as mass times acceleration, multiplying expression~(\ref{ro1})
by $ d \mathbf{r}$, and transforming the double vector product in the last term
using the Lagrange formula, we obtain the formula for elementary work:
    $$ 
d \left( \frac{m v^2}{2} \right) = - \frac{\partial U}{ \partial \mathbf{r}}
d \mathbf{r} +
d \left( \frac{m}{2} \left[ \mathbf{r}, \mathbf{\Omega} \right]^2 \right).
$$ 
    Therefore, the energy in the rotating reference frame is
    \begin{equation}    \label{ro4}
E = \frac{m v^2}{2} + U - \frac{m}{2} \left[ \mathbf{r}, \mathbf{\Omega}
\right]^2.
\end{equation}
    In the nonrelativistic case, the particle velocity in the nonrotating
reference frame is
    \begin{equation}    \label{ro5}
\mathbf{v_0} = \mathbf{v} + \left[ \mathbf{\Omega}, \mathbf{r} \right].
\end{equation}
    Substituting~(\ref{ro5}) in~(\ref{ro4}) and taking the nonrelativistic
expression $\mathbf{L} = m \left[ \mathbf{r} , \mathbf{v} \right]$
for the angular momentum into account, we obtain the equality
    $$ 
E = \frac{m v_0^2}{2} - \mathbf{\Omega} \cdot \mathbf{L} + U.
$$ 
    for the energy in the rotating reference frame.
        For a free particle, setting $U=0$ and choosing the rotation direction
according to formula~(\ref{v2}), i.e. $\mathbf{\Omega} = (0, 0, - \Omega)$,
we again obtain expression~(\ref{v10v}) for the energy:
    $$ 
E = \frac{m v_0^2}{2} + \Omega L_z .
$$ 
    Hence, the change in the particle energy measured as force times
translation in the nonrelativistic case results in the same formula~(\ref{v10v}).

    Obviously, the new value of particle energy~(\ref{v10v}) obtained in
different ways does not coincide with the energy value $E'$ measured in
the same flat space-time but in the nonrotating reference frame.
    At first glance, this contradicts the invariance of energy
expression~(\ref{NEnergyEgen}) under coordinate transformations.
    This apparent contradiction is resolved if in defining the energy in some
reference frame, we use the Killing vector $(1,0,0,0)$ corresponding to
time translations in that reference frame.
    Passing to another reference frame, we use another Killing vector
to define the energy.
    Therefore, if the Killing vector in the rotating reference frame
is $\zeta =(1,0,0,0)$, then it has the form $\zeta =(1,0,-\Omega,0)$
in the nonrotating reference frame.
    Conversely, the Killing vector describing time translations in
a nonrotating reference frame has the coordinates $\zeta' =(1,0,0,0)$
in that frame.
    In passing to the rotating frame, its coordinates
are $\zeta' =(1,0,\Omega,0)$.
    The vector $\zeta$ can be represented as a sum of the different Killing
vectors of the flat space-time.

    We note that in quantum field theory in a rotating reference frame,
an expression for the energy operator of the field quanta similar
to~(\ref{v10v}) appears.
    Hence, Eq.~(3) in~\cite{Vilenkin80} for the statistical operator
includes the expression $H - \mathbf{L} \mathbf{\Omega}$
(where $H$ is the Hamiltonian) as an energy operator in the rotating reference frame,
and this expression corresponds to~(\ref{v10v}).
    An expression similar to~(\ref{v10v}) was used in~\cite{Letaw80,Duffy2003}
to define the positive frequency modes in a rotating reference frame
(formula~(13) in~\cite{Letaw80} and formula~(2.4) in~\cite{Duffy2003}).

    The equations of motion for free particles in a rotating
reference frame are
    \begin{equation}
c^2 \frac{d t}{d \lambda } = E - \Omega L_z, \ \ \
\frac{d z}{d \lambda } = p_z = {\rm const} ,
\label{uv1}
\end{equation}
    \begin{equation}
r^2 \frac{d \varphi}{d \lambda } = \left( 1 - \frac{\Omega^2 r^2}{c^2}
\right) L_z + \frac{\Omega r^2}{c^2} E,
\label{uv2}
\end{equation}
    \begin{equation}
c^2 \left( \frac{d r}{d \lambda } \right)^2 = R_f, \ \ \ R_f =
(E - \Omega L_z)^2 - m^2c^4 - (p_z c)^2 - \frac{L^2_z c^2}{r^2} .
\label{uv3}
\end{equation}

\begin{figure}[th]
\centering
  \includegraphics[width=58mm]{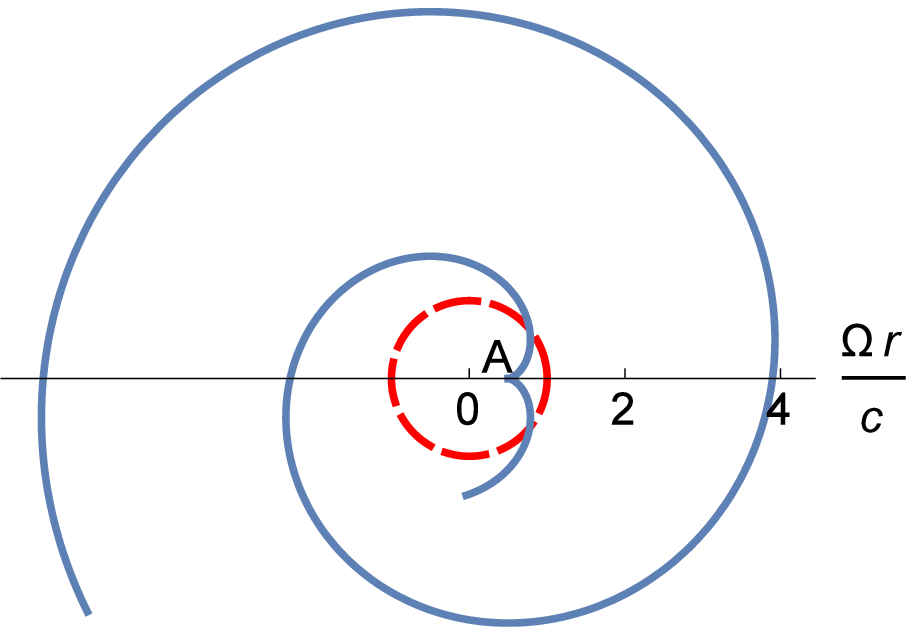} \ \ \ \ \
  \includegraphics[width=35mm]{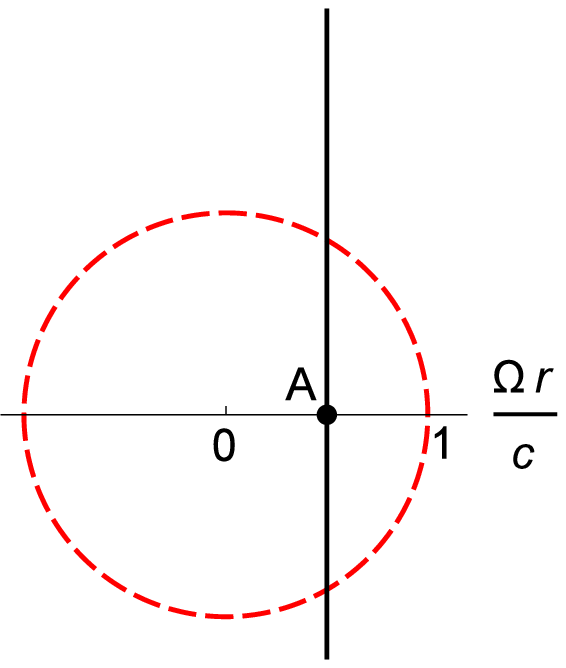}
\caption{The trajectory of a free particle with the minimum possible energy at
the point $A= ( c/ 2 \Omega,\, 0)$
in the rotating reference frame ($r, \varphi$), at the left,
and in the stationary reference frame ($r, \varphi'$), at the right.}
\label{MinEn}
\end{figure}

    For fixed values of $|p_z|$ and $r=r_A $ inside the static limit,
the possible values of the energy~$E^{(\zeta)}$ satisfy the inequality
    $$ 
E^{(\zeta)} \ge \sqrt{ m^{\mathstrut 2} c^4 +(p_z c)^2 }\,
\sqrt{1 - \frac{\Omega^2 r_A^2}{c^2} }.
$$ 
    The minimum of $E^{(\zeta)}$ is realized for particles with
    $$ 
        L_z = - \frac{\Omega r_A^2 }{c^2} \, E', \ \ \
E' = \sqrt{\frac{ m^{\mathstrut 2} c^4 +(p_z c)^2 }{
\displaystyle 1 - \frac{\Omega^2 r_A^2}{c^2}}}.
$$ 
    For such particles,
    $$ 
\frac{d \varphi}{ d t} = \Omega \left( 1- \frac{r_A^2}{r^2} \right).
$$ 
    The trajectory equation for such particles projected on the plane
$XOY$ can be found from formulas~(\ref{uv2}) and (\ref{uv3}):
    $$ 
\varphi - \varphi_A = \pm \left( \sqrt{ \frac{r^2}{r_A^2} -1} -
\arccos \left( \frac{r_A}{r} \right) \right).
$$ 
    The trajectory of such a particle for $r_A=c/(2 \Omega)$ and $\varphi_A=0$,
is shown in Fig.~\ref{MinEn}.
    In the stationary reference frame, these particles move with
the velocity projection on the plane $XOY$ equal to $r_A \Omega$
in the direction of rotation of the movable frame with the target
distance $r_A$ from the $z$ axis.

    We note that the expression
    $$ 
\frac{ E^{(\zeta)} }{m c^2} = \sqrt{1 - \frac{\Omega^2 r_A^2}{c^2} }.
$$ 
    for the minimum energy with $p_z=0$ corresponds to the value of
the radiation frequency shift
    $$ 
\frac{\nu}{\nu_0 } = \sqrt{1 - \frac{\Omega^2 r_A^2}{c^2} },
$$ 
under the transverse Doppler effect~\cite{LL_II}
for a body rotating with the angular velocity~$\Omega$ at the distance~$r_A$
from the rotation axis, which was experimentally confirmed in~\cite{Kundig63}.

    For the Killing vector of translation along the coordinate~$x^0=c t$,
we have the equality
    $$ 
(\zeta , \zeta)= 1 - \frac{\Omega^2 r^2}{c^2} .
$$ 
     Therefore, $\zeta$ becomes spacelike beyond the static limit,
and the particle energy~$E^{(\zeta)}$ beyond the static limit in the rotating
reference frame can take negative values with an absolute value as large
as needed (see problem 10.15 in~\cite{LPPT}).
    This case is analogous to the case of a rotating black hole, in
whose ergosphere the Killing vector of $t$-translation is spacelike
and the ``energy at infinity'' corresponding to this vector can be
negative.

    For a fixed value of $p_z$, we obtain the necessary condition for
a negative particle energy in a rotating reference frame beyond the static limit
from expressions~(\ref{uv1}) and (\ref{uv3}) and $dt / d \lambda > 0$:
    \begin{equation}    \label{Novu}
L_z < - \sqrt{\frac{p_z^2 c^2 + m^2 c^4}{ \Omega^2 - (c/r)^2 }}, \ \ \ \
r> c/\Omega .
\end{equation}
    If a particle with negative energy is at a distance $r$ from the origin,
then its velocity in the stationary reference frame $v$,
as follows from~(\ref{Novu}), satisfies the inequality
    \begin{equation}    \label{Novuu}
E < 0,  \ \ r > \frac{c}{\Omega} \ \ \ \Rightarrow
\ \ \ \frac{v}{c} > \frac{c}{r \Omega}.
\end{equation}
    States with zero energy are also possible beyond the static limit.
        Their properties were considered in detail in~\cite{GribPavlov2016d}.

\vspace{4mm}
{\centering \section{\large Penrose effect in rotating reference frames}
\label{secConl}}

    The existence of states with negative energy in the ergosphere of
a rotating black hole allowed Penrose~\cite{Penrose69} to propose a process
of energy extraction from a black hole that now bears his name.
    If a particle in the ergosphere decays into two particles with one
having negative energy, then the energy of the other particle is greater
than the energy of the decaying particle.
    If the second particle goes to spatial infinity, then energy is
thus extracted from the black hole.
    We note that the energy in this process means the energy related to
the time translation in the reference frame reducible to
the flat frame not rotating at the spatial infinity.
    We call this energy ``the energy at infinity''~\cite{MTW}.
    The corresponding Killing vector in the ergosphere is spacelike.

    In a rotating reference frame, the energy of particles beyond the
static limit can also be negative.
    As previously shown in~\cite{GribPavlov2016d}, such particles
cannot get inside the static limit, for instance, to the rotating Earth.
    Is it possible to obtain additional energy inside the static limit
of a rotating reference frame during the decay of a body into
two fragments?

    In~\cite{GribPavlov2016d}, we considered the case where
the energy gain was obvious, but it disappeared in the particular
experiment of measuring in the center of the rotating frame
because of the necessity to change the particle parameters
(decay fragments) for them to hit the point of measurement.
    Here, we consider an example where the Penrose effect in
a rotating frame turns out to be observable.

\begin{figure}[th]
\centering
  \includegraphics[width=100mm]{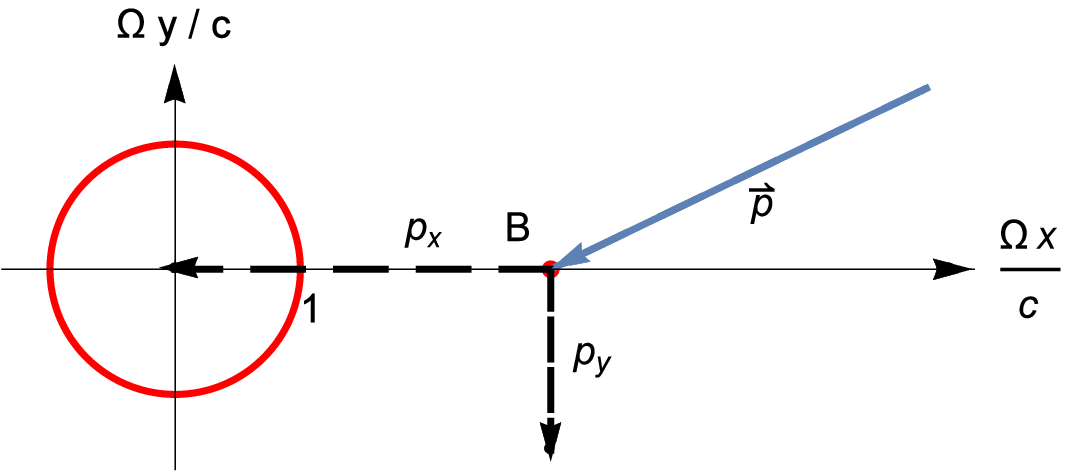}
\caption{Particle decay with an energy increase in a rotating reference frame.}
\label{Penr}
\end{figure}

    Let a particle of mass $m$ and energy~$E'$ move on
the plane $ (XOY) $ in a stationary Cartesian frame from infinity
to a point $B$ with the Cartesian coordinates $(x_B, 0,0)$, and
have some (negative) momentum projections $p_x$ and $p_y$.
    Let the moving particle decay into two particles of equal mass
at the point $B$: the first flies to the origin with the momentum $p_x$,
and the second moves vertically downward with the momentum $p_y$,
as shown in Fig.\ref{Penr}.
    From the energy-momentum conservation law, we obtain the values of
mass of the fragments
    \begin{equation}    \label{PR3}
\mu = \frac{\sqrt{m^{\mathstrut 4} c^4 - 4 p_x^2 p_y^2}}{2 E'}
\end{equation}
    and their energies in the stationary inertial reference frame
    \begin{equation}    \label{Er1}
E'_1 = \frac{m^2 c^4 + 2 (p_x c)^2 }{2 E'}, \ \ \
E'_2 = \frac{m^2 c^4 + 2 (p_y c)^2 }{2 E'}.
\end{equation}
    The energy of the original particle in the rotating frame is
    $$ 
E = E' + \Omega L_z = E' + \Omega ( x p_y - y p_x) =
c \sqrt{m^2 c^2 + p_x^2 + p_y^2} + \Omega x_B p_y,
$$ 
    and the energy of the first fragment is $E_1=E'_1$.

    The energy difference between the first fragment and the original
particle in the rotating frame is
    \begin{equation}    \label{DER}
E_1 - E = x_B \Omega |p_y| - \frac{m^2 c^4 + 2 (p_y c)^2 }{2 E'} = -E_2.
\end{equation}
    Obviously, for a sufficiently large $x_B$, this quantity is positive
(the energy of the second fragment is correspondingly negative), and we thus
obtain an energy gain similar to the Penrose process in rotating black holes.
    The energy of the first fragment flying to the region inside the static
limit, where the laboratory can be located and the energy in the rotating
frame can be measured, is greater than the energy of the original particle.

    The energy excess can obviously be as large as desired with a choice
of large~$x_B$.
    But the first particle energy value $E_1$ itself cannot exceed
the value $E'$ (see formula~(\ref{Er1})).
    Therefore, in a rotating reference frame, an unlimited energy gain
of the fragment and original particle occurs as a result of the negative
energy of the original particle. And it cannot reach the static limit.

    In the case of zero energy of the original particle, its momentum
projections are related by the condition
    $$ 
p_y = - \sqrt{ \frac{m^2 c^4 + p_x^2 c^2}{ \Omega^2 x_B^2 - c^2 } }.
$$ 
    In this case, the energy gain is
    $$ 
\frac{E_1 -E}{E'} = \frac{m^2 c^2 + 2 p_x^2}{ 2 (m^2 c^2 + p_x^2) }
\left( 1 - \frac{c^2}{x_B^2 \Omega^2} \right).
$$ 
    This quantity is always less than unity but can be close to it
for sufficiently large $|p_x|$.
    The energy of the second fragment in the rotating frame is $- E_1$,
which can be easily verified.
   Hence, the particle flying from spatial infinity with zero energy
and decaying outside the static limit gives two fragments of which
one has positive energy in the rotating frame and can be observed
inside the static limit and even at the rotation axis.

    In the case $p_y/p_x < c/ (\Omega x_B)$, the trajectory of
the original particle could pass inside the static limit
(if decay does not occur).
    In this case, the gain in the fragment energy in the rotating
frame can be measured by directly comparing the fragment energy and
the energy of the particle that is analogous to the original
particle but reaches the static limit without decaying.
    Let $p_y = \alpha p_x c/ (\Omega x_B)$.
        We assume that $|p_x| \gg mc $ and $ \alpha  \ll 1$.
        The energy gain of the original particle and the fragment flying
to the origin is then (according to formula~(\ref{DER}),
a quantity close to $\alpha c p_x$.
    The quantity $\alpha c$ corresponds to the rotation velocity of
the reference frame at such a target distance from the rotation axis
where the trajectory of the original particle in the stationary frame passes.
   Therefore, the energy increase in this case corresponds to
the value of the Doppler shift when the observer recedes from
    the radiation source with the velocity $\alpha c$.

     In the case of the Penrose effect for rotating black holes,
there are constraints on the value of the relative velocity of
the fragments~\cite{Chandrasekhar,BardeenPressTeukolsky72}.
     We therefore consider the question of the relative velocity of
the fragments in the presented example for the Penrose effect in
a rotating reference frame.
    The relative velocity $\mathbf{v}_{\rm rel}$ for two particles
with nonzero rest masses at the moment of the original particle decay can
be found by passing to the reference frame related to one of these particles.
    Hence, in the frame related to the first particle, the components of
the 4-velocity of particles $u_{(n)}$, $n=1,2$, are
    $$ 
u^i_{(1)} = (1,0,0,0), \ \ \ \ u^i_{(2)} =
\frac{ \displaystyle \left( 1, \frac{\mathbf{v}_{\rm rel}}{c}  \right)}{
 \displaystyle  \sqrt{1- \frac{v^{2}_{\rm rel}}{c^2}}}.
$$ 
    Therefore,
    $$ 
u^i_{(1)} u_{(2) i} = 1 \biggl/ \sqrt{ 1- \frac{v^{2}_{\rm rel}}{c^2} },
\ \ \ \ v_{\rm rel} = c\, \sqrt{1 - (u^i_{(1)} u_{(2) i})^{-2} } .
$$ 
    The last expression is independent of the choice of the reference frame
because the product $u^i_{(1)} u_{(2) i}$ is invariant.
    Using expressions~(\ref{PR3}) and (\ref{Er1}) in the considered case,
we obtain
    $$ 
\frac{v_{\rm rel}}{c} = \frac{\displaystyle   \sqrt{1 - 4 \frac{\mu^2}{m^2}
}}{ \displaystyle  1 - 2 \frac{ \mu^2 }{ m^2 } } =
\sqrt{
\frac{m^2 c^4}{E'_1 E'_2} \left( 1- \frac{m^2 c^4}{ 4 E'_1 E'_2} \right) }.
$$ 
    Because $\mu \le m/2$, we have $0\le v_{\rm rel}/c \le 1$.
        In this case, if $\mu =0$, then $v_{\rm rel}/c = 1$ and if $\mu =m/2$,
then $v_{\rm rel}/c = 0$,which is possible only if $p_x =p_y =0$.
    The smaller the value of the fragment masses~$\mu$,
the larger the relative velocity is.

    To realize the considered Penrose process, the condition $E_2<0$
must be satisfied, which by virtue of~(\ref{Novuu}) near the static limit
leads to the velocity of the second fragment $v_2 \to c$, and consequently
to the value of the relative velocity $v_{\rm rel} \to c$.
    Therefore, to realize the Penrose effect for decay near the static limit,
the relative velocities of fragments must be near the speed of light.

    If the decay is far from the static limit, $x_B \gg c/ \Omega$,
then to obtain the restrictions for the relative velocity of the fragments
in the case $E_2 < 0$, we use the inequality for the possible values
of~$p_y$ at the given~$E'$ following from equality~(\ref{DER}):
    $$ 
 -E' \frac{x_B \Omega }{c} -
\sqrt{ \left( E' \frac{x_B \Omega }{c} \right)^2 - 2 m^2 c^4 }  \le
2 c p_y \le  -E' \frac{x_B \Omega }{c} +
\sqrt{ \left( E' \frac{x_B \Omega }{c} \right)^2 - 2 m^2 c^4 } .
$$ 
    Hence,
    $$ 
p_y \le  - \frac{m^2 c^4}{2 E' x_B \Omega }.
$$ 
    The minimum value of the relative velocity is reached for
$p_y = - m c^2 /(2 \Omega x_B)$, $ p_x = 0$ and is equal to $c / \Omega x_B$.
    Therefore, for the relative velocity of the fragment in the Penrose
process during decay far from the static limit, we have the inequality
    \begin{equation}
\frac{v_{\rm rel}}{c} \ge \frac{c}{\Omega x_B} .
\label{otnmin}
\end{equation}
    Inequality~(\ref{otnmin}) shows which decay processes in
the inertial reference frame can be treated in terms of
the Penrose effect in a rotating reference frame.

\vspace{4mm}
{\centering \section{\large Conclusion}
\label{secEPenr}}

    In a uniformly rotating reference frame, there is a surface beyond which
no body can stay at rest in this frame.
  By analogy with a rotating black hole, we call this surface the static limit.
    Beyond this surface, there are states with negative energy defined in
the rotating frame.
    If a particle decays beyond the static limit, then the energy of a fragment
that comes inside the static limit and is registered by an observer can be
greater than the energy of the original decaying particle.
   This phenomenon is analogous to the Penrose effect for rotating black holes.

\vspace{5mm}
{\bf Acknowledgments.}\,
    This research is supported by the Russian Foundation for Basic research
(Grant No. 18-02-00461 a).
    The work of Yu.V.P. was supported by the Russian Government Program of
Competitive Growth of Kazan Federal University.

\vspace{2mm}


\begin{thebibliography}{99}
\setcounter{enumiv}{0}
\itemsep=1mm

\bibitem{Ehrenfest1909}
P. Ehrenfest,
``Gleichf\"{o}rmige Rotation starrer K\"{o}rper und Relativit\"{a}tstheorie'',
{{\it Phys.~Z.} {\bf 10}, 918 (1909)};
English transl.
``Uniform rotation of rigid bodies and the theory of relativity''
in {\it Relativity in Rotating Frames:
Relativistic Physics in Rotating Reference Frames},
eds. G.~Rizzi and M.\,L.~Ruggiero,
Kluwer Academic Publ., Boston (2004),
\href{http://dx.doi.org/10.1007/978-94-017-0528-8_1}
{pp.~3--4.}  

\bibitem{EhrenfestVrash}
G. Rizzi and M.\,L. Ruggiero, (eds.),
\href{http://dx.doi.org/10.1007/978-94-017-0528-8}
{\it Relativity in Rotating Frames: Relativistic Physics}
\href{http://dx.doi.org/10.1007/978-94-017-0528-8}
{\it in Rotating Reference Frames},
Kluwer Academic Publ., Boston (2004).

\bibitem{GribPavlov2016d}
A.\,A. Grib, Yu.\,V. Pavlov,
``Comparison of particle properties in Kerr metric and in rotating coordinates'',
\href{http://dx.doi.org/10.1007/s10714-017-2238-3}
{{\it Gen. Relativ. Gravit.} {\bf 49}, 78  (2017)}.

\bibitem{MTW}
C.\,W.~Misner, K.\,S.~Thorne and J.\,A.~Wheeler,
{\it Gravitation}, Freeman, San Francisco (1973).

\bibitem{Penrose69}
R. Penrose,
``Gravitational collapse: The role of general relativity'',
\href{http://adsabs.harvard.edu/abs/1969NCimR...1..252P}
{\it Rivista Nuovo}
\href{http://adsabs.harvard.edu/abs/1969NCimR...1..252P}
{{\it  Cimento, Num. Spec.} {\bf I},  252--276 (1969)}.

\bibitem{Unruh76}
W.\,G. Unruh,
``Notes on black-hole evaporation'',
\href{https://doi.org/10.1103/PhysRevD.14.870}
{Phys. Rev.~D. {\bf 14}, 870--892 (1976)}.

\bibitem{LL_II}
L.\,D.~Landau and E.\,M.~Lifshitz, {\it Field Theory} [in Russian], Nauka,
Moscow (1988); {\it The Classical Theory of Fields}, Pergamon, Oxford (1983).

\bibitem{Fok}
V.~Fock, {\it Theory of Space, Time and Gravitation} [in Russian], GIFML,
Moscow (1961); English transl., Pergamon, Oxford (1964).

\bibitem{EinsteinInfeld}
A. Einstein and L. Infeld {\it The Evolution of Physics. The Growth of
Ideas from Early Concepts to Relativity and Quanta},
Simon and Schuster, New York (1954).

\bibitem{Friedmann}
A.\,A. Friedman, {\it The World as Space and Time} [in Russian],
Academia, Petrograd (1923); Nauka, Moscow (1965);
English transl., Minkowski Inst. Press, Montreal (2014).

\bibitem{BornETO}
M. Born, {\it Einstein's Theory of Relativity}, Dover, New York (1962).

\bibitem{FizikaKosmosa}
R.\,A. Syunyaev, ed., {\it Physics of the Cosmos: Little Encyclopedia}
[in Russian], Soviet Encyclopedia, Moscow (1986).

\bibitem{BoyerLindquist67}
R.\,H. Boyer and R.\,W. Lindquist,
``Maximal analytic extension of the Kerr metric'',
\href{http://dx.doi.org/10.1063/1.1705193}
{{\it J.~Math. Phys.} {\bf 8}, 265--281 (1967)}. 

\bibitem{Chandrasekhar}
S.~Chandrasekhar, {\it The Mathematical Theory of Black Holes},
Clarendon, New York (1983).

\bibitem{LPPT}
A.\,P.~Lightman,  W.\,H.~Press,  R.\,H.~Price, and S.\,A.~Teukolsky,
{\it Problem Book in Relativity and Gravitation},
Princeton Univ. Press, Princeton, N. J. (1975).

\bibitem{LL_I}
L.\,D.~Landau and E.\,M.~Lifshitz, {\it Mechanics} [in Russian], Nauka,
Moscow (1988); English transl. prev. ed., Pergamon, Oxford (1976).

\bibitem{Vilenkin80}
A.~Vilenkin,
``Quantum field theory at finite temperature in a rotating system'',
\href{https://doi.org/10.1103/PhysRevD.21.2260}
{{\it Phys. Rev. D} {\bf 21}, 2260--2269 (1980)}.  

\bibitem{Letaw80}
J.\,R. Letaw and J.\,D. Pfautseh,
``Quantized scalar field in rotating coordinates'',
\href{https://doi.org/10.1103/PhysRevD.22.1345}
{{\it Phys. Rev. D} {\bf 22}, 1345--1351 (1980)}.  

\bibitem{Duffy2003}
G. Duffy and A.\,C. Ottewill,
``Rotating quantum thermal distribution'',
\href{https://doi.org/10.1103/PhysRevD.67.044002}
{{\it Phys. Rev. D} {\bf 67}, 044002 (2003)}.

\bibitem{Kundig63}
W.~K\"{u}ndig,
``Measurement of the transverse Doppler effect in an accelerated system'',
\href{https://doi.org/10.1103/PhysRev.129.2371}
{{\it Phys. Rev.} {\bf 129}, 2371--2375 (1963)}. 

\bibitem{BardeenPressTeukolsky72}
J.\,M. Bardeen, W.\,H. Press, and S.\,A. Teukolsky,
``Rotating black holes: locally nonrotating frames, energy extraction,
and scalar synchrotron radiation'',
\href{http://dx.doi.org/10.1086/151796}
{\it Astrophys. J.}
\href{http://dx.doi.org/10.1086/151796}
{{\bf 178}, 347--369 (1972)}.  

\end{thebibliography}
\end{document}